\documentclass[aps,prb,twocolumn,superscriptaddress,showpacs]{revtex4}
\usepackage{graphicx}
\usepackage{calc}
\usepackage{bm}

\begin{document}

\title{Experimental investigation of the edge states structure at fractional filling factors.}

\author{E.V.~Deviatov}
\email[Corresponding author. E-mail:~]{dev@issp.ac.ru}
 \affiliation{Institute of Solid State
Physics RAS, Chernogolovka, Moscow District 142432, Russia}

\author{V.T.~Dolgopolov}
\affiliation{Institute of Solid State Physics RAS, Chernogolovka,
Moscow District 142432, Russia}

\author{A.~Lorke}
\affiliation{Laboratorium f\"ur Festk\"orperphysik, Universit\"at
Duisburg-Essen, Lotharstr. 1, D-47048 Duisburg, Germany}

\author{W.~Wegscheider}
\address{Institut f\"ur Angewandte und Experimentelle Physik,
Universitat Regensburg, 93040 Regensburg,
 Germany}

\author{A.D.~Wieck}
\affiliation{Lehrstuhl f\"ur Angewandte Festk\"orperphysik,
Ruhr-Universit\"at Bochum, Universit\"atsstrasse 150, D-44780
Bochum, Germany}

\date{\today}

\begin{abstract}
We experimentally study electron transport between edge states in
the fractional quantum Hall effect regime. We find an anomalous
increase of the transport across the 2/3 incompressible fractional
stripe in comparison with theoretical predictions for the smooth
edge potential profile. We interpret our results as a first
experimental demonstration of the intrinsic structure of the
incompressible stripes arising at the sample edge in the
fractional quantum Hall effect regime.
\end{abstract}


\maketitle

The concept of the edge states (ES) was firstly introduced by
Halperin~\cite{halperin} to describe the transport phenomena in
two-dimensional (2D) systems in the integer quantum Hall effect
regime. ES, arising at the intersections of distinct Landau levels
with Fermi level, can be introduced for both
sharp~\cite{buttiker} and smooth~\cite{shklovsky} edge potential
profile. Experimentally the existence of ES was proved not only in
transport experiments along the sample edge (for a review see
Ref.~\onlinecite{haug}) but also across
it~\cite{alida,rdiff,relax}.

This single-electron description is not applicable to the
fractional quantum Hall effect, which is fundamentally a many-body
phenomenon~\cite{laughlin}. Electron system forms a many-body
ground state below the Fermi level and an excited state above it.
A set of compressible stripes, separated by the incompressible
regions with fractional fillings, is expected to exist at the
sample edge for the case of the smooth edge
potential~\cite{Beenakker}. One-dimensional chiral Luttinger
liquid states are predicted theoretically for  the opposite case
of the sharp potential jump at the sample
edge~\cite{macdonald,wenPRB41}. The ES structure in the later case
was found to be determined by the hierarchical
structure~\cite{haldane} of the bulk ground
state~\cite{macdonald,wenPRB41}.   The transport along the sample
edge, however, is not sensitive to the form of the edge potential,
but only to the filling factor in the bulk. It can be described by
modified Buttiker formulas~\cite{Beenakker,macdonald} in good
agreement with experiments~\cite{crossgates} in Hall-bar geometry
with cross-gates. For this reason, these experiments can not be
used to distinguish between the proposed models. In real
experiments the strength of the potential profile can not be
regarded as infinitely large, so the model of the smooth edge
potential seems to be more realistic. On the other hand,
experiments on tunnelling into the fractional edge demonstrate
 the complicated structure of edge excitation
spectrum~\cite{chang,milliken} which is expected for the  sharp
edge~\cite{wenPRB41}. This controversial situation demand the
investigation of the fractional ES structure for real samples.

\begin{figure}
\includegraphics*[width=0.6\columnwidth]{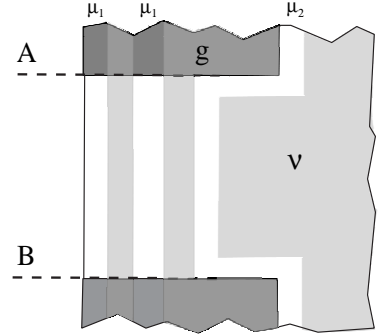}%
\caption{ Schematic diagram of the gate-gap region in the
pseudo-Corbino sample geometry.  The dark area represents the
Schottky-gate. The light gray area indicates incompressible
regions in the sample. In a quantizing magnetic field at total
filling factor $\nu$, one set of the edge states (the number is
equal to the filling factor under the gate, $g$; $g<\nu$) is
propagating under the gate along the etched edge of the sample and
carry the electrochemical potentials $\mu_1$. The other edge
states (their number is $\nu-g$) are going along the gate edge and
carry electrochemical potentials $\mu_2$. In the gate-gap region,
both sets of the edge states are running in parallel, leading to
the current across the incompressible region between them, if
$\mu_1 \neq \mu_2$.\label{sample}}
\end{figure}

It was shown theoretically~\cite{chamon} that after smoothening
the sharp edge potential, a transition takes place and new
branches of ES appears. The edge that had one right-moving ES
before the transition, for example, is having two right-moving ES
and one left-moving ES. The same prediction about the edge
reconstruction with smoothing the edge potential was made by using
the composite-fermion language~\cite{chklovskyCF}. Experimentally,
the edge reconstruction picture can be verified by studying  the
electron transport across the sample edge in the quasi-Corbino
geometry, because  this experiment was shown to be very sensitive
to the ES structure~\cite{topdefect}.

Here, we experimentally study  electron transport between
different ES in the fractional quantum Hall effect regime. We find
an anomalous increase of the transport, at some filling factors,
in comparison with the prediction of the simple Beenakker
model~\cite{Beenakker} of fractional ES. We interpret our results
as a first experimental demonstration of  the intrinsic structure
of the incompressible stripes arising at the reconstructed sample
edge in the fractional quantum Hall effect regime in accordance
with the model of Wen and Chamon~\cite{chamon}.

Our samples are fabricated from two molecular beam epitaxial-grown
GaAs/AlGaAs heterostructures with different carrier concentrations
and mobilities. One of them (A) contains a 2DEG located 210~nm
below the surface.  The mobility at 4K is 1.93 $\cdot
10^{6}$cm$^{2}$/Vs and the carrier density 1.61 $\cdot 10^{11}
 $cm$^{-2}$.
For heterostructure B the  corresponding parameters are 150~nm,
1.83 $\cdot 10^{6}$cm$^{2}$/Vs and 8.49 $\cdot 10^{10}$cm$^{-2}$.

Measurements are performed in the quasi-Corbino sample
geometry~\cite{alida,rdiff}. In this geometry a sample has two
non-connected etched mesa edges (the inner and the outer ones,
like Corbino disks) with independent ohmic contacts at every edge.
ES, originating from one mesa edge, are redirected to the other
mesa edge by using split-gate technique. As a result, ES from
independent ohmic contacts run together along the outer etched
edge of the sample in the gate-gap region as depicted in
Fig.~\ref{sample}. The gate-gap width (AB) is 5~$\mu$m for samples
from wafer A and 0.5~$\mu$m for ones from wafer B. The available
fractional filling factors and the electron concentration in the
ungated region were obtained from usual magnetoresistance
measurements. Also, magnetocapacitance measurements were performed
to characterize   the electron system under the gate. The contact
resistance at low temperature is about 100 $\Omega$ per contact,
as was determined from 2-point magnetoresistance measurements. The
temperature of the experiment is 80~mK, the magnetic field is up
to 14~T.

We study the $I-V$ characteristics of the gate-gap region by
applying dc voltage between the outer and the inner ohmic contacts
and by measuring the appeared dc current. In the integer quantum
Hall effect regime, the dissipative conductance component is close
to zero in the two-dimensional electron gas (2DEG). For this
reason, the measuring current is the current between two groups of
independently-contacted ES in the gate-gap, see Fig.~\ref{sample}.
If the equilibration length for transport between them is smaller,
than the gate-gap width, we can expect a full equilibration in the
gate-gap and a linear $I-V$ trace. In the opposite regime, charge
transfer does not change the chemical potentials of ES
significantly and the applied voltage $V$ directly affects on the
value of potential barrier between ES, leading to it's
disappearance at some positive voltage $V_{th}>0$ (because of the
negative electron charge). Zero potential barrier, in order, means
zero equilibration length between ES. Thus, in spite of the
strongly non-linear $I-V$ trace in this case, a positive branch
above $V_{th}$ have to be linear like in the opposite regime. It
was experimentally established~\cite{alida,rdiff}, that
 $V_{th}$ and the slope of the linear part of the positive
branch are universal characteristics, reflecting the potential
barrier value between ES and the redistribution of the
electrochemical potential imbalance between them. They coincide
with theoretical values (the spectral gap and the equilibrium
redistribution, obtained from Buttiker formulas~\cite{buttiker})
with a possible 10\% deviation. This 10\% deviation is connected
to the potential disorder at the sample edge and is a constant for
the given sample. It does not depend on the cooling procedure and
other occasional parameters.

\begin{figure}
\includegraphics[width= 0.8\columnwidth]{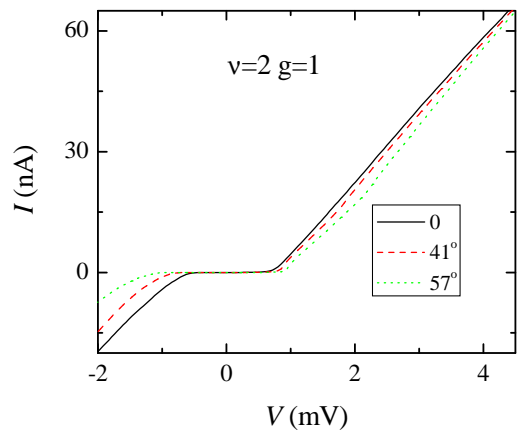}%
\caption{  $I-V$ curves for the sample A for filling factors
$\nu=2$ and $g=1$ at different tilt angles: $\theta=0$ (solid
line), $\theta=41^\circ$ (dashed line), $\theta=57^\circ$ (dotted
line). Experimental slope of the positive branch is constant and
equals to $2.2 h/e^2$, the equilibrium Buttiker value is $2
h/e^2$. Perpendicular magnetic field equals to 3.34~T, gate
voltage $V_g=268~mV$. \label{IV21}}
\end{figure}

Typical $I-V$ curves in the integer quantum Hall effect regime are
shown in Fig.~\ref{IV21} for the filling factor combination
$\nu=2, g=1$. The $I-V$ traces reflect electron transport between
two spin-split edge states, because at $\nu=2$ two spin-split
energy levels are filled in the bulk. The equilibration length in
this case can reach a millimeter~\cite{mueller}, which is much
higher than the gate-gap width for both samples A and B. Every
experimental $I-V$ trace is strongly non-linear, with the linear
part on the positive branch. Tilting the sample plane with respect
to the magnetic field allows us to introduce an in-plane field
component, keeping the filling factor by adjusting the value of
total field. As it can be seen from Fig.~\ref{IV21}, in-plane
field affects on the linear part of the $I-V$ only by increasing
$V_{th}$ value, leaving the slope to be non-affected. We can
conclude for our samples that the in-plane magnetic field does not
change the equilibrium mixing of ES in the integer quantum Hall
effect regime.

\begin{figure}
\includegraphics[width= 0.9\columnwidth]{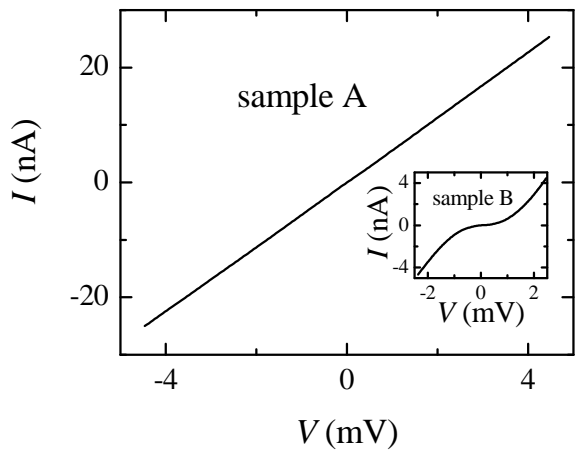}%
\caption{  $I-V$ curves for the samples A (main field) and B
(inset) for filling factors $\nu=2/3$ and $g=1/3$. Magnetic field
equals to 10~T for the sample A and to 4.85~T for the sample B.
The gate-gap width and the temperature of the experiment are
different for both samples: 5~$\mu$m and 80~mK for the sample A
and 0.5~$\mu$m and 30~mK for the sample B. The experimental slope
of the linear $I-V$ curve is $6.8 h/e^2$, the equilibrium Buttiker
value is $6 h/e^2$. \label{IV23131}}
\end{figure}

Examples of the $I-V$ curves for fractional fillings are shown in
Fig.~\ref{IV23131} for $\nu=2/3, g=1/3$. As it can be seen from
the inset to the figure, $I-V$ curve is strongly non-linear at the
lowest temperature of 30~mK for the sample B  with the smallest
gate-gap width 0.5~$\mu$m (also non-linear $I-V$'s for fractional
fillings were reported in Ref.~\onlinecite{alida2}). By increasing
the temperature and the gate-gap width (80~mK and 5~$\mu$m for the
sample A) the equilibration length between fractional ES can be
made smaller than the gate-gap width, leading to the fully linear
$I-V$, see the main part of Fig.~\ref{IV23131}. In this Letter we
put our attention on the analysis of the linear $I-V$ curves at
fractional filling factors.

\begin{figure}
\includegraphics[width= 0.85\columnwidth]{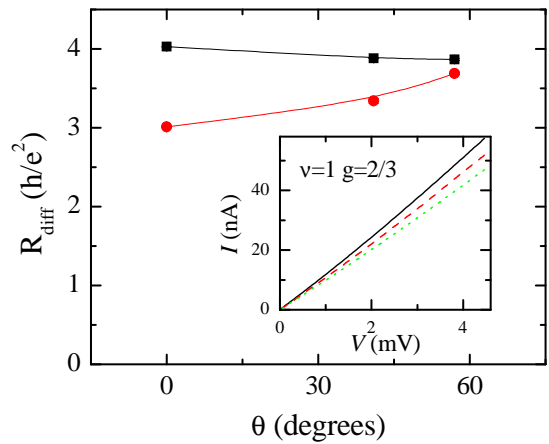}%
\caption{  Experimental slopes of linear $I-V$ curves for the
sample A for filling factor combinations $\nu=1$ and $g=1/3$
(squares) and $\nu=1$ and $g=2/3$ (circles) as functions of tilt
angle. Error bars are within the symbol size. The equilibrium
Buttiker value is $4.5 h/e^2$ for both filling factor
combinations. The normal magnetic field is constant and equals to
6.68~T Inset shows the original $I-V$ curves for filling factors
$\nu=1$ and $g=2/3$ at different tilt angles: $\theta=0$ (solid
line), $\theta=41^\circ$ (dashed line), $\theta=57^\circ$ (dotted
line). $I-V$ curves are independent of the cooling cycle and well
reproducible. \label{rdiff}}
\end{figure}

The most intriguing results are obtained for filling factor
combinations $\nu=1, g=1/3$ and $\nu=1, g=2/3$. In
Fig.~\ref{rdiff} experimental slopes of linear I-V curves are
shown in dependence on the sample tilt angle in magnetic field.
The slopes for $\nu=1, g=1/3$ behaves like in the integer case:
they are practically independent of the in-plane field.
Experimental values for $\nu=1, g=2/3$ differ significantly from
ones for $\nu=1, g=1/3$ in normal field and are approaching to
them with increasing the in-plane field component, see
Fig.~\ref{rdiff}. It can be also seen  from the inset to
Fig.~\ref{rdiff}, where the original $I-V$ curves for different
tilt angles are shown for filling factor combination $\nu=1,
g=2/3$. Let us stress that $I-V$ curves are  well reproducible.

Because the gate-gap region in our quasi-Corbino geometry is
formed electrostatically by using the split-gate, it is obvious to
use the Beenakker model~\cite{Beenakker} of fractional ES in
smooth edge profile to describe the experiment. In this model the
edge potential is supposed to be smooth enough to introduce the
local filling factor $\nu_c$. At the sample edge it is
monotonically changing from the bulk value $\nu=1$ to zero.
Incompressible stripes are formed around fractional local filling
factors $\nu_c=2/3$ and 1/3. Buttiker formulas can easily be
generalized to this situation~\cite{Beenakker}:
$$
I_\alpha=\frac{e}{h}\nu_\alpha\mu_\alpha - \frac{e}{h}\sum_\beta
T_{\alpha\beta}\mu_\beta,
$$
where $I_\alpha$ is the current in ES $\alpha$, corresponding to
the fractional filling factor $\nu_\alpha$ and connected to a
contact with electrochemical potential $\mu_\alpha$;  $e,h$ are
the electron charge and the Plank constant repectively,
 $T_{\alpha\beta}$ are the  Buttiker coefficients for the transmission from
contact $\beta$ to contact $\alpha$.

This formulas can easily be applied to our experimental geometry,
while the only difference from the integer case is the presence of
constant weight coefficients $\nu_\alpha$. Slopes of the linear
$I-V$ curves can be calculated in the assumption of full
equilibration between all ES in the gate-gap:
$$
R_{diff}=\frac{h}{e^2}\frac{\nu}{g(\nu-g)}
$$
Correspondingly, we can expect that like for integer ES (i) linear
$I-V$ curve means the full equilibration between ES in the
gate-gap (ii) experimental slopes should coincide with calculated
ones within 10\%, as it was discussed above;  (iii) these slopes
should be independent of the in-plain magnetic field component,
like presented in Fig.~\ref{IV21}. Moreover, we can expect from
the calculation that $I-V$ slopes for the filling factor
combinations $\nu=1,g=1/3$ and $\nu=1,g=2/3$ will coincide
exactly. In the experiment, however, the former filling factor
combination (as well as $\nu=2/3,g=1/3$) behaves as described,
while the $I-V$ slope for $\nu=1,g=2/3$ is in 1.5 times smaller
than the theoretical value, approaching the values for
$\nu=1,g=1/3$ with increasing the in-plane field component. Thus,
we can conclude that electron transport at $\nu=1,g=2/3$ is
anomalously enhanced in comparison with the equilibrium
electrochemical potential redistribution, while at $\nu=1,g=1/3$
it is about the theoretical value. For these two filling factor
combinations the filling factor in the gate-gap $\nu=1$ is the
same as well as other parameters of the sample edge (potential
profile, disorder, etc). The only difference is the incompressible
stripe, which separates ES from inner and outer contacts in the
gate-gap: it correspond to $\nu_c=2/3$ for $\nu=1,g=2/3$ and to
$\nu_c=1/3$ for $\nu=1,g=1/3$.

\begin{figure}[t]
\includegraphics[width= 0.85\columnwidth]{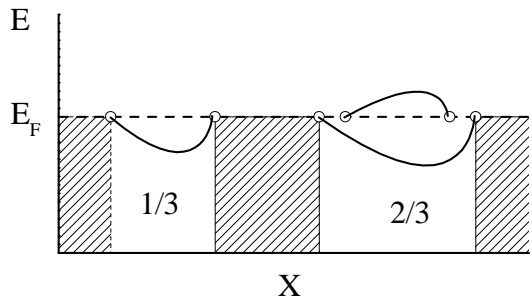}%
\caption{  Schematic energy diagram of the sample edge in the
fractional quantum Hall effect regime. Hatched regions represents
compressible stripes with electrons at the Fermi level. In the
incompressible "puddles" between them the energy of the fractional
ground state is sketched (it is asymmetric because of the edge
potential). It is a simple Laughlin fractional ground state for
the puddle with 1/3 local filling factor. The ground state for the
2/3 local filling is more complicated: it is electron ground state
for the filling factor 1 and hole fractional for the filling
factor 1/3. It leads to the two counterpropagating branches of ES
per edge of this puddle. ES are denoted by open circles.
\label{fracES}}
\end{figure}

This behavior can not be explained within the model of Beenakker,
where the local filling factor $\nu_c=2/3$ has no difference from
any other one. On the other hand, 2/3 has very special character
in the model of sharp edge potential profile of
MacDonald~\cite{macdonald}. Here $\nu=2/3$ is regarding as the
electron ground state of the filling factor 1 and the Laughlin
hole fractional one with positive fractional charge 1/3. Both ones
give their contributions into the ES formation, leading to two
counterpropagating ES at one edge: the outer integer for electrons
and the inner fractional for holes. This model can not be directly
applied to our experiment, because the electrostatical edges in
any case are not sharp and even etched ones are very doubt. It was
predicted~\cite{chamon,chklovskyCF} that while smoothing the edge
profile, edge reconstruction occurs and quantum Hall "puddles" are
forming with local fractional filling factors, see
Fig.~\ref{fracES}.  Each boundary of the fractional quantum Hall
puddle is still can be regarded as a sharp boundary of quantum
Hall system with particular fractional filling factor. This leads
to the formation of a number of counterpropagating fractional ES
at every sample edge. Of course, the net current along the edge is
still depend on the bulk filling factor only, so in our experiment
the detailed structure of ES is important only in the gate-gap,
where the charge transfer across the edge occurs. Also, the etched
edge seems to be sharp enough to apply this model of reconstructed
ES.

Let us consider the filling factor combination $\nu=1,g=2/3$. At
low temperature the bulk of the sample is in the incompressible
state at filling factor $\nu=1$ in the ungated region and at
$g=2/3$ under the gate. Approaching the etched edge,
incompressible "puddles" of lower fractional fillings are formed,
see Fig.~\ref{fracES}. In the gate-gap they correspond to $\nu_c =
2/3; 1/3$, while only $g_c=1/3$ is present under  the gate. It is
clear, that incompressible puddle with $\nu_c=2/3$ in the gate-gap
is directly connected to the incompressible state $g=2/3$ under
the gate, while the puddles $\nu_c=1/3$ and $g_c=1/3$ forms the
incompressible stripe along the etched edge like it shown in
Fig.~\ref{sample}. It means that the picture of compressible and
incompressible states, presented in Fig.~\ref{sample} still
survive in the fractional quantum Hall regime, but the structure
of ES is very different. Fractional ES are formed at the edges of
every incompressible puddle. The current across the sample edge
can flow only by tunnelling between these ES through the
incompressible regions and by diffusion in the compressible ones.
At the filling factor combination $\nu=1,g=2/3$ the tunnelling in
the gate-gap takes place across the 2/3 incompressible puddle, see
Figs.~\ref{sample},\ref{fracES}. As it is described above,
fractional ES at every edge of the puddle with $\nu_c=2/3$ are the
counterpropagating  electron integer ES with current
$\mu_\alpha\frac{e}{h}$ and  hole fractional with current
$-\frac{1}{3}\mu_\alpha\frac{e}{h}$, leading to the sum current
$\frac{2}{3}\mu_\alpha\frac{e}{h}$ per one edge, see
Fig.~\ref{fracES}. Because of the complex nature of the ES for
$\nu_c=2/3$, they are not far away from each other and we can
expect that only these ES are mixing their electrochemical
potentials in the gate-gap. A simple calculation gives in this
case the resistance of $3 h/e^2$. It is in 1.5 times smaller than
it would be if all ES in the gate-gap mixed their electrochemical
potentials and is in fact observing in the experiment in normal
magnetic field, see Fig.~\ref{rdiff}. The in-plane magnetic field
increases the fractional gaps (it was verified for our samples by
usual magnetocapacitance spectroscopy), transforming fractional
quantum Hall puddles into the  stripes of significant width. It
makes the proposed mechanism to be ineffective and the only way is
to mix the electrochemical potentials of all present ES in the
gate-gap, like in Beenakker model. As a result, the differential
resistance increases to the value of $9/2$ $h/e^2$. As about the
other fillings under consideration, $\nu=1,g=1/3$ and
$\nu=2/3,g=1/3$, tunnelling should occur between ES in the
$\nu_c=1/3$ quantum Hall puddle. There is no complex ES structure
in this case and ES are far away from each other. The proposed
 mechanism is ineffective and mixing between all existing ES
in the gate-gap takes place at any in-plane field, as we observe
in the experiment, see Figs.~\ref{IV23131},\ref{rdiff}.

As a result, we study electron transport across the sample edge in
the fractional quantum Hall effect regime in the quasi-Corbino
sample geometry. At the filling factor combination $\nu=1,g=2/3$
we observe an anomalous increasing of the  current in comparison
with the prediction of the simple Beenakker model~\cite{Beenakker}
of fractional ES. We interpret our results as a first experimental
demonstration of  the intrinsic structure of the incompressible
stripes arising at the reconstructed sample edge in the fractional
quantum Hall effect regime, in accordance with the model of Wen
and Chamon~\cite{chamon}.

We wish to thank  D.E.~Feldman for fruitful discussions and
A.A.~Shashkin for help during the experiment. We gratefully
acknowledge financial support by the RFBR, RAS, the Programme "The
State Support of Leading Scientific Schools", Deutsche
Forschungsgemeinschaft, and SPP "Quantum Hall Systems", under
grant LO 705/1-2. E.V.D. acknowledges support by Russian Science
Support Foundation.

\end{document}